\documentclass[fdp,fleqn]{w-art}
\usepackage{times}
\usepackage{w-thm}
\usepackage[dvips]{graphicx}
\usepackage{epsf}                        
\usepackage{amssymb}
\usepackage{latexsym}                    
\usepackage{amsfonts}                    

\begin{document}
\DOIsuffix{theDOIsuffix}
\Volume{55}
\Issue{1}
\Month{01}
\Year{2007}
\pagespan{3}{}





\title[Charged Particle-like Branes in ABJM: A Summary]{Charged Particle-like Branes in ABJM: A Summary}


\author[N. Gutierrez]{Norberto Guti\'errez\inst{1,}
  \footnote{E-mail:~\textsf{norberto@string1.ciencias.uniovi.es$\;$ --$\;$ Speaker}}}
\address[\inst{1}]{Department of Physics,  University of Oviedo, Avda.~Calvo Sotelo 18, 33007 Oviedo, Spain}

\author[Y. Lozano]{Yolanda Lozano\inst{1,}\footnote{E-mail:~\textsf{ylozano@uniovi.es}}}

\author[D. Rodriguez-Gomez]{Diego Rodr\'{\i}guez-G\'omez\inst{2,}\footnote{E-mail:~\textsf{drodrigu@physics.technion.ac.il}}}
\address[\inst{2}]{Department of Physics, Technion, Haifa, 3200, Israel}
\address[\inst{2}]{Department of Mathematics and PhysicsUniversity of Haifa at Oranim, Tivon, 36006, Israel}


\begin{abstract}

We study the effect of adding lower dimensional brane charges to the 't Hooft monopole, di-baryon and baryon vertex configurations in $AdS_4 \times \mathbb{P}^3$. We show that these configurations capture the background fluxes in a way that depends on the induced charges, requiring additional fundamental strings to cancel the worldvolume tadpoles. The dynamics reveal that the charges must lie inside some interval, a situation familiar from the baryon vertex in $AdS_5 \times S^5$ with charges. For the baryon vertex and the di-baryon the number of fundamental strings must also lie inside an allowed interval. Some ideas about the existence of these bounds in relation to the stringy exclusion principle are given.

\end{abstract}
\maketitle                   





\section{Introduction}

On the field theory side of the $AdS/CFT$ correspondence, there are operators whose scaling dimension is of order $N$, typically corresponding to non-perturbative states in the gravity side realized as branes wrapping calibrated cycles. In the context of the more familiar $AdS_5/CFT_4$ correspondence, a canonical example is the di-baryon \cite{GK,Berenstein:2002ke}, which corresponds to a wrapped D3-brane, in addition to the baryon vertex \cite{Witten}. The latter corresponds to a D5-brane wrapping the whole internal space. Since it captures the $N$ units of 5-form flux, it requires $N$ fundamental strings ending on it. These strings can be thought of as external quarks (or Wilson lines in the fundamental representation) and this naturally suggests to identify this wrapped D5-brane with the $\epsilon$ tensor of the $SU(N)$ gauge group of the field theory.

The $AdS_4/CFT_3$ case is less understood in general. The field theory side is provided by the ABJM model, conjectured to describe M2 branes probing a $\mathbb{C}^4/\mathbb{Z}_k$ singularity \cite{ABJM}. Their theory is an ${\cal N}=6$ quiver Chern-Simons-matter theory with gauge group $U(N)_k\times U(N)_{-k}$, where $(k, -k)$ are the CS levels of each gauge group. The theory has a marginal superpotential, as well as an appropriate large $k$ limit $N^{1/5}<<k<<N$ that allows for a weak coupling regime. In turn, in the gravity dual, the appropriate description is in terms of Type IIA string theory on $AdS_4\times\mathbb{P}^3$, with various RR fluxes. In this background, various particle-like branes were already discussed in \cite{ABJM}. The $\mathbb{P}^3$ space has $H^{q}(\mathbb{P}^3)=\mathbb{R}$ for even $q\le 6$. Thus, it is possible to have $D0$, $D2$, $D4$ and $D6$ branes wrapping a topologically non-trivial cycle \cite{ABJM, Gaiotto:2009mv}. While the $D2$ and $D6$ branes capture RR flux and develop worldvolume tadpoles, the $D0$ and $D4$ branes do not, and therefore should correspond to gauge-invariant operators. It has been argued that the D0-brane corresponds to a di-monopole operator in the CFT, while the D4-brane should correspond to a di-baryon-like one. The D6-brane, very much like the baryon vertex in $AdS_5$, requires $N$ fundamental string ending on it. Its dual operator should then naturally involve the $\epsilon$ tensor of the gauge theory. On the other hand, the D2-brane wrapped on the $\mathbb{P}^1\subset \mathbb{P}^3$ develops a tadpole that has to be cancelled with $k$ fundamental strings. The dual operator is a monopole 't Hooft operator, realized as a ${\rm Sym}_k$ product of Wilson lines \cite{ABJM}.

These gravitational configurations admit a natural generalization by allowing non-trivial worldvolume gauge fluxes. It is the aim of this paper to generalize the spectroscopy of wrapped branes by adding such non-trivial worldvolume gauge fields. These generalized configurations are of potential interest for some $AdS/CMT$ applications, for example as candidates for holographic anyons in ABJM, as discussed recently in \cite{Hartnoll:2006zb, KL}.

Allowing for a non-trivial worldvolume gauge field has the effect of adding lower dimensional brane charges, and in some cases, additional fundamental strings will be required to cancel the worldvolume tadpoles. From that point of view, the generalized configurations are similar to holographic Wilson loops. We will see that the D2 and D6-branes do not differ much from the zero charge case, although they are stable only if the induced charges lie below some upper bound.
This situation is familiar from the baryon vertex with magnetic flux in $AdS_5\times S^5$, studied in \cite{JLR}. In these cases the energy of the bound state increases with the charge that is being induced. However adding charges allows to construct more general baryon vertex configurations. We will see that for the D6-brane the number of quarks that forms the bound state can be increased in this manner. In turn, the D4-brane with flux behaves quite differently from the fluxless case, since it will require fundamental strings ending on it as opposed to the vanishing worldvolume flux case. As we will see, the study of its dynamics reveals that the whole configuration is stable if the magnetic flux lies within a given interval, being maximally stable for an intermediate value, and reducing to free quarks at the boundaries.

\section{Particle-like branes in $AdS_4\times \mathbb{P}^3$ with magnetic flux}

In this section we generalize the particle-like brane configurations in \cite{ABJM} to include a non-vanishing magnetic flux, using the conventions taken in \cite{GLR} for the $AdS_4\times \mathbb{P}^3$ background.

\subsection{The di-baryon vertex}

This brane wraps the $\mathbb{P}^2$ in $\mathbb{P}^3$ and, since it does not capture any background fluxes, it does not require any fundamental strings ending on it. On the other hand, the $\mathbb{P}^2$ is not a spin manifold, and the D4-brane should carry a half-integer worldvolume gauge field flux through the $\mathbb{P}^1\subset\mathbb{P}^2$ due to the Freed-Witten anomaly \cite{Freed:1999vc}. This half-integer worldvolume flux can be cancelled through a shift of $B_2$. This motivated \cite{AHHO} to include a flat $B_2=-2\pi\, J$ field in the dual IIA background.

We can now consider a more general configuration where we add extra worldvolume flux $F={\cal N}\,J$, such that the total worldvolume flux is $F_T=(\mathcal{N}+1)\, J$, with $\mathcal{N}\in 2\mathbb{Z}$ (being $\mathcal{N}=0$ the minimal case). So we have $\mathcal{F}=2\pi\, F$, that is, the $B_2$ shift and the extra half unit of worldvolume flux cancel each other and we can effectively work as if we had no background $B_2$-field and $F=\mathcal{N}\, J$. In this case the D4-brane mass satisfies $m_{D4} L=N+k\,{\cal N}^2/8$. For vanishing flux the mass is proportional to N, thus naturally admitting an interpretation as a di-baryon. In addition, the D4-brane with magnetic flux captures the $F_2$ background flux and D0-brane charge through the couplings:
\begin{equation}
S_{CS}=k\,\frac{{\cal N}}{2}\, T_{F1} \int dt A_t\, , \ \ \ \ S_{CS}=\frac{{\cal N}^2}{8}\, T_0 \int_{\mathbb{R}} C_1\,.
\end{equation}
Therefore $k\,{\cal N}/2$ fundamental strings are required to end on it in order to cancel the tadpole. On the other hand, the mass can be interpreted as the energy of a threshold BPS intersection of ${\cal N}^2/8$ D0-branes and a D4-brane. Notwithstanding, as noted in \cite{AHHO}, there are relevant higher curvature corrections to be considered \cite{Green:1996dd}. In this case, the total D0 charge gets modified to $\mathcal{N}^2/8-1/24$. We should note however that if we want to study the dynamics of the D4-brane with fundamental strings attached in the probe brane approximation, we need to take the strings distributed uniformly on the D4. Therefore, all supersymmetries will be broken. Nevertheless, since both the wrapped cycle and the worldvolume flux are topologically non-trivial, we expect the system to be at least perturbatively stable. The same reasoning holds for the other configurations.

A discussion about the possible role played by the B-field in the quantization of Abelian fields in $AdS$ can be found in \cite{GLR}. As there, here we will simply assume that suitable boundary conditions are chosen allowing for the corresponding wrapped objects and, as we have done for the D4-brane, we will include the effect of the (flat) $B$-field. The D4-brane with zero flux would be identified with the di-baryon operator in the CFT side. Now, since once the worldvolume flux is turned on extra F1 strings are required, we should expect such dual operator to involve $n_f=k\,\mathcal{N}/2$ Wilson lines in the fundamental representation of $U(N)\times U(N)$.

\subsection{The 't Hooft monopole}

Let us now consider the D2-brane wrapping the $\mathbb{P}^1$ in $\mathbb{P}^3$, which require $q=k$ fundamental strings attached. This brane has been identified in \cite{ABJM} with a ('t Hooft) monopole operator. Given that it wraps a spin manifold, it does not capture the Freed Witten anomaly, and as such, the quantization condition for $F_T$ \cite{Bachas:2000ik} is $\frac{1}{2\pi}\int F_T=\frac{1}{2}\,(\mathcal{N}+1)\in \mathbb{Z}$, with minimal flux $\mathcal{N}=-1$. In this case, there is D0-brane charge induced in the configuration, since
\begin{equation}
\label{D0_in_D2}
S_{CS}=2\pi\, T_2\int_{\mathbb{R}\times \mathbb{P}^1} C_1\wedge F=\frac{{\cal N}}{2}\, T_0 \int_{\mathbb{R}} C_1\, .
\end{equation}
We see that our system is actually formed by a $D2-D0$ bound state, which hints to a non-supersymmetric configuration. For this reason conjecturing a dual operator seems a much harder task.

\subsection{The baryon vertex}

Finally, the D6-brane wrapping the whole $\mathbb{P}^3$ is the analogue of the baryon vertex in $AdS_5 \times S^5$ \cite{Witten}, and requires the addition of $q=N$ fundamental strings.

Let us now switch on a gauge flux $F_T={\cal N}J$.\footnote{Although this represents a slight change in the conventions compared to the previous sections.} Since $\mathbb{P}^3$ is spin, the appropriate quantization condition is $\frac{1}{2\pi}\int F_T=\frac{{\cal N}}{2} \in \mathbb{Z}$. As in \cite{AHHO}, the D6 brane turns out to be sensitive to higher order curvature couplings, although, in the end, the relevant CS terms are
\begin{equation}
S_{CS}=\frac16 (2\pi)^2\, T_6 \int_{\mathbb{R}\times \mathbb{P}^3} P[F_2]\wedge F_T \wedge \Bigl(2\pi F_T +3\, P[B_2]\Bigr)\wedge A=
k\, \frac{{\cal N}({\cal N}-2)}{8}\, T_{F1} \int dt A_t\, .
\end{equation}
Therefore, due to the magnetic flux introduced, the number of fundamental strings that must end on the D6 is $q=N+k\,{\cal N}({\cal N}-2)/8$ (note that this is always an integer). On the other hand, the magnetic flux and the $B_2$ field also induce D2 and D0-brane charge in the configuration in a consistent way (\cite{GLR}).

\section{Study of the dynamics: Charge bounds}

In this section we study the stability in the $AdS$ direction of the brane configurations that we have previously discussed. We follow the calculations in \cite{BISY} and \cite{Maldacena} (see also \cite{JLR} for similar results for the baryon vertex with magnetic flux in $AdS_5\times S^5$). In order to analyze the stability in the radial $\rho$-direction (for details on specific conventions we refer to \cite{GLR}), we have to consider both the D$p$-brane wrapped on the $\mathbb{P}^{p/2}$ and the $q$ fundamental strings stretching between the brane and the boundary of $AdS$. Following the analysis in \cite{BISY} the equations of motion come in two sets: the bulk equation of motion for the strings, and the boundary equation of motion, which contains as well a term coming from the D$p$-brane. Both equations can be combined into just\footnote{We have parameterized the worldvolume coordinates by $(t,x)$ and the position in $AdS$ by $\rho=\rho(x)$.}

\begin{equation}
\label{eqmotion}
\frac{\rho^4}{\sqrt{\frac{16\rho^4}{L^4}+\rho'^2}}=\frac{1}{4}\,\beta\, \rho_0^2\, L^2\, ,
\end{equation}
where $\rho_0$ is the position of the brane in the holographic direction, and $\beta\in [0,1]$ is defined by
\begin{equation}
\label{beta2}
\beta=\sqrt{1-\Bigl(\frac{2Q_p}{L\,q\,T_{F1}}\Bigr)^2}\, , \ \ \ \ Q_p=\frac{\pi^{p/2}\,T_p}{(\frac{p}{2})!\, g_s}(L^4 + (2\pi{\cal N})^2)^{p/4}\,.
\end{equation}
Since $\beta$ is a function of the magnetic flux the square root imposes a bound on its possible values, and therefore it restricts the possible charges that can be dissolved in the D$p$-brane.

By integrating the equation of motion, we find that the size of the configuration is given by \linebreak ${\ell}=L^2/4\rho_0\int_1^\infty dz \; \beta/(z^2\sqrt{z^4-\beta^2})$, where $z=\rho/\rho_0$. This expression has the same form as that found in \cite{BISY,JLR,Maldacena,RY}, and can also be solved in terms of hypergeometric functions. Note that the dependence on the location of the configuration $\rho_0$, and on $L^2$, is also the same, which is again a non-trivial prediction of the AdS/CFT correspondence for the strong coupling behavior of the gauge theory.

The binding energy of the configuration can be obtained from the total on-shell energy, by subtracting the (divergent) energy of its constituents. When the D$p$-brane is located at $\rho_0=0$ the strings stretched between 0 and $\infty$ become radial, and therefore correspond to free quarks. At this location the energy of the D$p$ vanishes. Therefore, the binding energy is given by:
\begin{equation}
\label{Ebinding}
E_{\rm bin}= q\, T_{F_1}\,\rho_0\Big(\sqrt{1-\beta^2}+\int_1^{\infty}dz\Big[\frac{z^2}{\sqrt{z^4-\beta^2}}-1\Big]-1\Big)\,.
\end{equation}
This expression has again the same form than the corresponding expressions in \cite{BISY,JLR,Maldacena,RY}.\footnote{In this case we have added the on-shell energy of the D$p$-brane.} For the values allowed for $\beta$ the binding energy per string is negative and decreases monotonically with it. Therefore, the configuration is stable for these values. When the D$p$-brane is charged the configuration with free quarks, corresponding to $\beta=0$, is degenerate, since it can be realized either as free radial strings stretching from 0 to $\infty$ plus a charged D$p$-brane, located at $\rho=0$, or as free radial strings stretching from  $\rho_0$ to $\infty$ plus a D$p$-brane located at $\rho_0$. Note that the location of the D$p$-brane has become a moduli of the system, and in both cases, since the strings are radial, the size of the configuration vanishes.

On the other hand, as we can also write $E_{bin}=-f(\beta)(g_s N)^{2/5}/\ell$ with $f(\beta)\ge 0$, we have that $dE/d{\ell}\ge 0$ and the configuration is stable. The behavior of $E_{\rm bin}$ as a function of the 't Hooft coupling and the size of the configuration is the same as in $AdS_5\times S^5$ \cite{Maldacena,RY,BISY}. The fact that it goes as $1/{\ell}$ is dictated by conformal invariance, while the behavior with $\sqrt{\lambda}$ is a non-trivial prediction for the non-perturbative regime of the gauge theory. Note that the same non-analytic behavior with $\lambda$ is predicted for ${\cal N}=4$ SYM in 3+1 dimensions and ABJM \cite{DPY,CW,RSY}. In fact, perturbative calculations can explain this behavior when extrapolated to strong coupling (\cite{Suyama}). The exact interpolating function was obtained in \cite{MP} for 1/6 and 1/2 BPS Wilson loops using topological strings.

It is possible to check that when the number of strings does not depend on $\beta$, i.e. for the 't Hooft monopole case, the configuration becomes more stable as $\beta$ increases. For studying the di-baryon and baryon vertex configurations it is necessary to use the specific dependence $q(\cal{N})$. Bounds for the allowed number of strings and magnetic flux can be obtained from a detailed dynamical study. In the following, we will summarize the results for the different configurations (\cite{GLR}).

In the t'Hooft monopole the minimum binding energy occurs for zero ${\cal N}$, for which  $\beta=\sqrt{1-1/(4\pi^2)}$, and $\beta=0$ is reached for ${\cal N}_{\rm max}/L^2=  \sqrt{1-1/(4\pi^2)}$, where stability breaks down. As a function of the 't Hooft coupling ${\cal N}_{\rm max}=\sqrt{8\lambda\,(4\pi^2-1)}$, which is exactly the same behavior that was encountered in \cite{JLR} for the baryon vertex in $AdS_5\times S^5$. This dependence on $\sqrt{\lambda}$ will hold for our di-baryon and baryon vertex configurations.

For the di-baryon $\beta_{max}=\sqrt{1-1/(4\pi^2)}$, although the number of strings depends now on the magnetic flux, and the minimum energy turns to occur for ${\cal N}=1.00 \, L^2$. The allowed values for the magnetic flux are those for which $1-\beta_{max} \le {\cal N}/L^2 \le 1+\beta_{max}$. On the other hand, the allowed interval for the number of fundamental strings ending on the D4-brane is given by $1-\beta_{max} \le q/(2\pi\, \sqrt{2kN}) \le 1+\beta_{max}$.

Finally, in the baryon vertex $\beta$ decreases monotonically with ${\cal N}$ until it reaches its minimum value $\beta=0$. The allowed values of the magnetic flux are ${\cal N}/L^2 \lesssim \sqrt{36\pi^2-1}/(2\pi)$, and the minimum energy configuration occurs for ${\cal N}/L^2 \sim 2.01$. It is however possible to find more general baryon vertex configurations with $q<N$ quarks if $N-q$ strings stretch between $\rho_0$ and 0. A similar analysis to the one in \cite{BISY} shows that
the boundary equation has to be modified, and the number of quarks can satisfy $1/2 (N+k{\cal N}({\cal N}-2)/8)(1+\sqrt{1-\beta^2})\le q \le N+k{\cal N}({\cal N}-2)/8$. 

For all the bounds presented here, the quarks are free for the minimum and maximum numbers allowed, where the configurations cease to be stable.

\section{Conclusions}

We have analyzed various configurations of particle-like branes in ABJM, focusing on the study of their dynamics. This study has revealed that new and more general monopole, di-baryon and baryon vertex configurations can be constructed if the particle-like branes carry lower dimensional brane charges.

We have seen that a new di-baryon configuration with external quarks can be constructed. In the presence of a non-trivial magnetic flux $F={\cal N}J$, with $J$ the K\"ahler form of the $\mathbb{P}^3$, this brane develops a tadpole that has to be cancelled with $k\,{\cal N}/2$ fundamental strings. A dynamical study reveals that the configuration is stable for $1-\sqrt{1-1/(4\pi^2)}\leq {\cal N}/L^2\leq 1+\sqrt{1-1/(4\pi^2)}$. It is perhaps significant that the value of the magnetic flux for which the configuration is maximally stable is that for which the (off-shell) energy of the ${\cal N}^2/8$ D0-branes dissolved in the D4-brane equals the (off-shell) energy of the D4-brane. This seems to point at some kind of degeneracy for the ground state.
The D2 and D6-brane (monopole and baryon vertex) configurations exist already for vanishing magnetic flux. Consistently, no minimum value is found in the study of their dynamics. 
In these cases the effect of the magnetic flux is to allow the construction of more general monopole and baryon vertex configurations. The simplest case is the D2-brane, for which the charge of the tadpole is not modified by the magnetic flux and the number of attached F-strings is still $k$. We have seen that the configuration formed by the bound state D2-D0 plus the $k$ F-strings is stable for  ${\cal N}/L^2 \leq \sqrt{1-\frac{1}{4\pi^2}}$, reducing to $k$ free quarks plus a D2-brane with $\frac{L^2}{2}\,\sqrt{1-\frac{1}{4\pi^2}}$ D0-brane charge
when the upper bound is reached. The D6-brane in turn is the analogue of the baryon vertex in $AdS_5\times S^5$ \cite{Witten}. The generalization of the later to include a non-vanishing magnetic flux was studied in \cite{JLR}. In that reference it was found that the magnetic flux had to be bounded from above in order to find a stable configuration, like for the D2 and D6 branes considered in this paper.
For the D6-brane the number of F-strings depends as well on the magnetic flux, but this fact does not modify substantially its dynamics. 

As we have mentioned, all the configurations that we have considered reduce to free quarks when the magnetic flux reaches the highest possible value (also the lowest for the D4-brane). For this value the brane can be located at an arbitrary position in $AdS$, with the free radial strings stretching from there to $\infty$. This is different from the free quark configuration of  \cite{BISY}, where the D5-brane is located at $\rho_0=0$, where it has zero-energy, and the radial strings stretch from 0 to $\infty$. For the maximum (and minimum, if applicable) value of the magnetic flux the D-brane is located at an arbitrary $\rho_0$, where it has some energy which is compensated by the lower energy of the strings stretching between $\rho_0$ and $\infty$. In the presence of magnetic flux the location of the D$p$-brane has therefore become a moduli of the system.

We have already stressed that in the probe brane approximation used all supersymmetries are broken. However, in analogy with the baryon vertex in $AdS_5\times S^5$ we expect that, at least when the charged branes are supersymmetric, some supersymmetries will be preserved when the strings join the brane at the same point. In this case we would have to consider the full DBI problem and look for spiky solutions.

It is significant that for all the configurations that we have discussed the binding energy is non-analytic in the 't Hooft coupling, with this non-analyticity being of the precise form of a square-root branch cut, like in $AdS_5\times S^5$. This hints at some kind of universal behavior based on the conformal symmetry of the gauge theory. 

An important question that remains open is what are the field theory realizations of the D$p$-branes with charge that we have considered. Since we do not expect that the D2 and D6 brane configurations are supersymmetric it is hard to have an intuition about the interpretation of the new charges in the field theory side. It is interesting to note that the number of extra fundamental strings required to cancel the worldvolume tadpole is that required to cancel the tadpole on the dissolved D2 branes. This might suggest that the dual operators are doped versions of the original ones with an operator representing the D2 branes. It is hard to be more precise, in particular due to the expected lack of SUSY. However, for the D4-brane with D0-charge one can expect that a supersymmetric spiky solution exists, in which case it makes sense to try to interpret the bounds on the magnetic flux in the gauge theory dual.  In field theory language the bound in $\beta$ would read:
\begin{equation}
\label{boundCFT}
N+\frac{{\cal N}^2}{8}\, k \le 2\pi\, n_f \sqrt{2\lambda}\, ,
\end{equation}
 where $n_f$ is the number of external quarks, which is a function of the magnetic flux: $n_f=k\,{\cal N}/2$, and $\lambda$ is the 't Hooft coupling. Therefore, at strong 't Hooft coupling we expect a bound on the baryon charge of (generalized) di-baryon configurations with $n_f$ external quarks. This should be related in some way to the stringy exclusion principle of \cite{MS}, although we have not been able to find a direct interpretation. Note that for all branes the bound on the magnetic flux exhibits the same non-analytic behavior with $\lambda$ as the binding energy, which seems to indicate that the bounds should have its origin in the conformal symmetry of the gauge theory.

\begin{acknowledgement}

The work of N.G. was supported by a FPU-MICINN Fellowship from the Spanish Ministry of Education and the European Social Fund. This work has been partially suported by the research grants MICINN-09-FPA2009-07122 and MEC-DGI-CSD2007-00042.

\end{acknowledgement}

\end{document}